\documentclass[12pt]{article}
\usepackage{latexsym,amsmath}
\usepackage{graphicx}

\begin{document}


\title{Phases of 4D Scalar-tensor  black holes  coupled to Born-Infeld non-linear
electrodynamics}

\author{Ivan Zh. Stefanov$^{1}$\thanks{E-mail: zhivkov@phys.uni-sofia.bg}\,,\:  Stoytcho S. Yazadjiev$^{1,2}$ \thanks{E-mail: yazad@phys.uni-sofia.bg}
\,\,\\{\footnotesize  ${}^{1}$Dept. of Theoretical Physics,
                Faculty of Physics}\\ {\footnotesize St.Kliment Ohridski University of Sofia}\\
{\footnotesize  5, James Bourchier Blvd., 1164 Sofia, Bulgaria }\\\\[-3.mm]
      {  \footnotesize ${}^{2}$Institut f\"{u}r Theoretische Physik,
         Universit\"{a}t G\"{o}ttingen} \\   {\footnotesize  Friedrich-Hund-Platz 1, D-37077 G\"{o}ttingen, Germany}\\\\[-3.mm]
 Michail D.~Todorov\thanks{E-mail: mtod@tu-sofia.bg}
\\ [-1.mm]{\footnotesize
{Faculty of Applied Mathematics and Computer Science}}\\
[-1.mm] {\footnotesize {Technical University of Sofia}}\\
[-1.mm] {\footnotesize 8, Kliment Ohridski Blvd., 1000 Sofia, Bulgaria}}

\date{}

\maketitle

\begin{abstract}
Recent results show that when non-linear electrodynamics is considered the no-scalar-hair theorems
in the scalar-tensor theories (STT) of gravity,
which are valid for the cases of neutral black holes and charged black holes in the Maxwell electrodynamics,
can be circumvented \cite{SYT1,SYT2}. What is even more, in the present work, we find new non-unique,
numerical solutions describing charged black holes coupled to non-linear electrodynamics in a special class of
scalar-tensor theories. One of the phases has a trivial scalar field and coincides with the
corresponding solution in General Relativity. The other phases that we find are characterized by the value of
the scalar field charge. The causal structure and some aspects of the stability of the solutions have also been
studied. For the scalar-tensor theories considered, the black holes have a single, non-degenerate horizon, i.e.,
their causal structure resembles that of the Schwarzschild black hole. The thermodynamic analysis of the stability
of the solutions indicates that a phase transition may occur.
\end{abstract}


\sloppy

\section{Introduction}

Among the most natural generalizations of General Relativity (GR) are the scalar-tensor theories of gravity
in which a single or multiple fundamental scalar fields are present as a possible remnant of a fundamental unified
theory like string theory or higher dimensional gravity theories \cite{DP}.
Different modifications of
scalar-tensor theories are attracting much interest also in cosmology and astrophysics.

Possible deviations from GR, especially the existence of new effects, due to the presence of the scalar field
would be of considerable interest. According to the no-scalar-hair conjecture, the black-hole solutions in the STT coincide with the solutions from GR. No-scalar-hair theorems treating the cases of static, spherically
symmetric, asymptotically flat, electrically neutral black holes
and charged black holes in the Maxwell electrodynamics have been proved for a large class of scalar-tensor theories
\cite{Be,Saa,BSen}. The scalar field in these cases is constant, and thus trivial, if one demands that the essential
singularity at the center of symmetry is hidden in a regular event horizon.

There are no no-scalar-hair theorems, however, in the non-linear
electrodynamics (NLED). It was first introduced by Born and Infeld
in 1934 to obtain finite energy density model for the electron
\cite{BI}. For more information on recent interests in non-linear
lagrangians of electrodynamics please see \cite{L}, also
\cite{Demianski}--\cite{Iran} and references cited therein for
gravitational aspects of NLED.

In the case of NLED, the energy-momentum tensor of the electromagnetic field
has a non-vanishing trace which is non-trivially coupled to the scalar field.
Hence, the electro-magnetic field acts as a
source of the scalar field and allows the existence of asymptotically flat, hairy black holes. Such solutions have
been found recently in different non-linear electrodynamics \cite{SYT1,SYT2}.
These solutions are hairy in a sense that the scalar field is not trivial. That hair, however, is
secondary since the solutions are determined uniquely by the values of their magnetic charge and mass,
and the value of the scalar field at infinity.
For other solutions describing asymptotically flat black holes with scalar hair (in which, however, the potential of the scalar field is not
positively semi-definite) we refer the reader to  \cite{NS}.

Another interesting effect that
occurs when NLED is considered is the presence of multiple black-hole phases in a certain class of STT.
Presence of non-unique solutions in STT has been found by
Damour and Esposito-Far\`{e}se in their works on neutron stars in the STT. One of the most interesting effects that
occur in these solutions is the spontaneous scalarization, a scalar field analogue of the spontaneous magnetization
of ferromagnets \cite{DF1,DF2,SS}. In the present work we consider the same STT and find multiple-phase black-hole
solutions numerically. Unlike the situation in \cite{SYT1,SYT2}, they are not
determined in a unique way by the values of their magnetic charge and mass and the value of the
scalar field at infinity.
The phase diagram of the present solutions is reminiscent of the phase transition between caged black holes and black strings in
higher dimensions \cite{Kol, Harmark}. The preliminary analysis gives us a reason to suppose that in the system we study a phase transition might occur.
The analysis of the thermodynamics of the solutions, however, is not sufficient to make a conclusion about the stability of the solutions
and about the existence of phase transitions. The full examination of the problem requires a perturbative analysis.


\section{Basic equations and qualitative investigation}

The theory we consider is presented \cite{SYT1, SYT2} both in the Jordan and in the Einstein frames. In the Einstein frame the action of the theory we consider is
\begin{eqnarray}\label{EFA}
S= {1\over 16\pi G_{*}}\int d^4x \sqrt{-g} \left[{\cal R} -
2g^{\mu\nu}\partial_{\mu}\varphi \partial_{\nu}\varphi -
4V(\varphi)+4{\cal A}^4(\varphi) L(X,Y)\right]. \nonumber
\end{eqnarray}
$V(\varphi)$ is the potential of the scalar field, $L(X)$ is the Lagrangian of the electromagnetic field,
${\cal A}^4(\varphi)$ is the function which determines the coupling between the electromagnetic field and the scalar field. We also have that
\begin{equation}
X = {{\cal A}^{-4}(\varphi)\over 4} F_{\mu\nu}{g}^{\mu\alpha} {g}^{\nu\beta} F_{\alpha\beta}, \label{X}  \,\,\,
Y = {{\cal A}^{-4}(\varphi)\over 4}  F_{\mu\nu}\left({ \star} F\right)^{\mu\nu}
\end{equation}
where ``$\star$''  stands for the Hodge dual with respect to the Einstein frame metric $g_{\mu\nu}$.
The type of the STT is determined by the explicit choice of the functions $V(\varphi)$ and ${\cal A}^4(\varphi)$.
The field equations obtained from action (\ref{EFA}) take the following form
\begin{eqnarray}
&&{\cal R}_{\mu\nu} = 2\partial_{\mu}\varphi \partial_{\nu}\varphi +  2V(\varphi)g_{\mu\nu} -
 2\partial_{X} L(X, Y) \left(F_{\mu\beta}F_{\nu}^{\beta} -
{1\over 2}g_{\mu\nu}F_{\alpha\beta}F^{\alpha\beta} \right)  \nonumber \\
&&-2{\cal A}^{4}(\varphi)\left[L(X,Y) -  Y\partial_{Y}L(X, Y) \right] g_{\mu\nu}, \nonumber  \\
&&\nabla_{\mu} \left[\partial_{X}L(X, Y) F^{\mu\nu} + \partial_{Y}L(X, Y) (\star F)^{\mu\nu} \right] = 0 \label{F},\\
&&\nabla_{\mu}\nabla^{\mu} \varphi = {dV(\varphi)\over d\varphi } -
4\alpha(\varphi){\cal A}^{4}(\varphi) \left[L(X,Y) -  X\partial_{X}L(X,Y) -  Y\partial_{Y}L(X, Y) \right], \nonumber
\end{eqnarray}
where $\alpha(\varphi) = {d{\ln \cal A}(\varphi)\over d\varphi}$.

In what follows the truncated\footnote{Here we consider the pure
magnetic case for which $Y=0$.  } Born-Infeld  electrodynamics
described by the Lagrangian

\begin{equation}
L_{BI}(X) = 2b \left( 1- \sqrt{1+ \frac{X}{b}} \right)\label{LBI}
\end{equation}

 will be considered. And $V(\varphi)$ will be equal to zero.


The anzats for metric of a static, spherically symmetric
space-time can be taken in the following form
\begin{equation}
ds^2 = g_{\mu\nu}dx^{\mu}dx^{\nu} = - f(r)e^{-2\delta(r)}dt^2 + {dr^2\over f(r) } +
r^2\left(d\theta^2 + \sin^2\theta d\phi^2 \right).
\end{equation}
Since the Born-Infeld NLED is invariant under electric-magnetic duality rotations we will study only the magnetically
charged black holes for which the electromagnetic field is given by
\begin{equation}
F = P \sin\theta d\theta \wedge d\phi
\end{equation}
and the magnetic charge is denoted by $P$.

The field equations reduce to the following coupled system of
ordinary differential equations:
\begin{eqnarray}
&&\frac{d\delta}{dr}=-r\left(\frac{d\varphi}{dr} \right)^2\label{EQDelta},\\
&&\frac{d m}{dr}=r^2\left[\frac{1}{2}f\left(\frac{d\varphi}{dr} \right)^2 - {\cal A}^{4} (\varphi)L(X)  \right] \label{EQm},\\
&&\frac{d }{dr}\left( r^{2}f\frac{d\varphi }{dr} \right)=
r^{2}\left\{-4\alpha(\varphi){\cal A}^{4}(\varphi) \left[L(X) -  X\partial_{X}L(X)\right] -
r f\left(\frac{d\varphi}{dr} \right)^3    \right\} \label{EQPhi}  ,
\end{eqnarray}
where
$$
f=1-\frac{2m}{r}
$$
and $ X $ reduces to:
\begin{equation}
X = {{\cal A}^{-4}(\varphi)\over 2} \frac{P^2}{r^4}.
\end{equation}
We will be searching for solutions which have a regular horizon
on which the scalar field $\varphi$ does not diverge. The
regularity of the transition between the Einstein and the Jordan
conformal frames requires the following restrictions on the
coupling function $0<{\cal A}(\varphi)<\infty$ for $r \geq r_{H}
$, where $r_{H}$ is the radius of the horizon, and we will
consider STT for which it is satisfied. The diversity and the
properties of the solutions depend strongly on the choice of the
functions ${\cal A}(\varphi)$ (respectively on $\alpha(\varphi)$).
Even when the functions  ${\cal A}(\varphi)$ are chosen to satisfy
the experimental constraints (see, for example, \cite{DF2}) we are
left with infinitely  many possibilities. That is why some
restrictions on the functions ${\cal A}(\varphi)$ representing the
STT should be imposed . In the present work we will consider only
theories for which $\varphi\,\alpha(\varphi)$ is non-negative or
non-positive for all values of $\varphi$. As we prove below,
theories for which $\varphi\,\alpha(\varphi)\ge 0$ for all values
of $\varphi$ do not admit (asymptotically flat) black hole
solutions with non-trivial scalar field. In theories with
$\varphi\,\alpha(\varphi)\le 0$ black holes (if they exist) have a
single non-degenerate event horizon when $\alpha(\varphi)=0$
admits only the solution $\varphi=0$.

Using the following equation
\begin{equation}
\frac{d }{dr}\left( e^{-\delta}r^{2}f\frac{d\varphi }{dr} \right)=4 r^2 e^{-\delta} \alpha(\varphi)
{\cal A}^{4}(\varphi)\left[X\partial_{X}L(X)- L(X)\right],\label{phianl}
\end{equation}
which is another form of equation (\ref{EQPhi}) and the fact that for the Born-Infeld Lagrangian (\ref{LBI}) holds
\begin{equation}
 X\partial_{X}L(X)- L(X)>0\label{EDHAM}
\end{equation}
we can draw some conclusions about the general properties of the solutions.

Let us multiply equation (\ref{phianl}) by $\varphi$ and then integrate it in the interval $r\in[r_{H},\infty)$ where we denote
the radius of the outer horizon (the event horizon) with $r_{H}$
\begin{align}
&\int\limits_{r_{H}}^{\infty}\varphi\, \frac{d }{dr}\left( e^{-\delta}r^{2}f\frac{d\varphi }{dr}\right) dr \notag \\
&\hspace{2cm} =4 \int\limits_{r_{H}}^{\infty}r^2 e^{-\delta} \varphi\, \alpha(\varphi)
{\cal A}^{4}(\varphi)\left[X\partial_{X}L(X)- L(X)\right] dr,
\end{align}
and after integrating by parts we get
\begin{multline}
 \lim_{r \to \infty}\left( \varphi\, e^{-\delta}r^{2}f\frac{d\varphi }{dr} \right) -
\left.\left(\varphi\, e^{-\delta}r^{2}f\frac{d\varphi }{dr} \right) \right|_{r=r_H}-
\int\limits_{r_{H}}^{\infty} \left[e^{-\delta}r^{2} f\left(\frac{d \varphi}{dr}\right)^{2}\right] dr \\
=\lim_{r \to \infty} {\cal{D}}\varphi  -
\int\limits_{r_{H}}^{\infty} \left[e^{-\delta}r^{2}f\left(\frac{d\varphi }{dr}\right)^{2}\right] dr \\
=4 \int\limits_{r_{H}}^{\infty}r^2 e^{-\delta} \varphi\, \alpha(\varphi)
{\cal A}^{4}(\varphi)\left[X\partial_{X}L(X)- L(X)\right] dr.
\end{multline}
In the second line we take advantage of the fact that we are looking for asymptotically flat,
black hole solutions
so $f(r_{H})=0$, $\lim_{r \to \infty}f(r)=1$, $\lim_{r \to \infty} \delta(r)=0$ and for the asymptotic value of the scalar field we impose
$\lim_{r\to \infty} \varphi(r)=\varphi_{\infty}$, where $\varphi_\infty=0$. The constant ${\cal{D}}$ denotes the scalar charge
which is defined as
\begin{equation}
{\cal{D}} = - \lim_{r \to \infty}  r^2 \frac{d\varphi}{dr}. \label{dilatoncharge}
\end{equation}
Since $\varphi$ is vanishing at infinity we finally get
\begin{eqnarray}
&&-\int\limits_{r_{H}}^{\infty} \left[e^{-\delta}r^{2}f\left(\frac{d\varphi }{dr}\right)^{2}\right] dr=\nonumber \\
&&\hspace{1cm}=4 \int\limits_{r_{H}}^{\infty}r^2 e^{-\delta} \varphi\, \alpha(\varphi)
{\cal A}^{4}(\varphi)\left[X\partial_{X}L(X)- L(X)\right] dr. \label{integ}
\end{eqnarray}
The total sign of the left-hand side of (\ref{integ}) is negative for black hole solutions with nontrivial $\varphi$. The sign of the integral on the right-hand side
depends on the sign of $\varphi\,\alpha(\varphi)$. If $\varphi\,\alpha(\varphi)\ge 0$ a contradiction is reached
so the assumption for the existence of asymptotically flat black holes with non-trivial scalar field in this theory is wrong.
Asymptotically flat
black-hole solutions with nontrivial scalar field exist  in scalar-tensor theories for which $\varphi\,\alpha(\varphi)\le 0$. Through a similar
examination we prove that these black holes have a single, non-degenerate horizon. Let us admit that more than one horizon exists.
Then we multiply equation (\ref{phianl}) by $\varphi$ again and integrate it by parts
in the interval $r\in[r_{-},r_{+}]$ where we denote the first inner horizon and the outermost, non-degenerate horizon with $r_{-}$
and $r_{+}$, respectively
\begin{eqnarray}
&&\left. \left( \varphi\, e^{-\delta}r^{2}f\frac{d\varphi }{dr} \right) \right|_{r=r_{+}}-
\left.\left(\varphi\, e^{-\delta}r^{2}f\frac{d\varphi }{dr} \right) \right|_{r=r_{-}}-
\int\limits_{r_{-}}^{r_{+}} \left[e^{-\delta}r^{2}f\left(\frac{d\varphi }{dr}\right)^{2}\right] dr \nonumber \\
&&\hspace{1cm}=4 \int\limits_{r_{-}}^{r_{+}}r^2 e^{-\delta} \varphi\, \alpha(\varphi)
{\cal A}^{4}(\varphi)\left[X\partial_{X}L(X)- L(X)\right] dr<0.
\end{eqnarray}
Having in mind that $f(r_{-})=0=f(r_{+})$ and that $f<0$ in the interval $r\in[r_{-},r_{+}]$
we reach a contradiction, which means that the
admission is incorrect.

Now, we only have to prove the non-existence of extremal black holes (black holes with degenerate event horizon).
For the scalar-tensor theories we consider $\alpha(\varphi)$ turns to zero only when $\varphi=0$.
Let us admit that an extremal black hole
with non-trivial scalar field exists. In this case, the left-hand side of (\ref{phianl}) is equal to zero. The right-hand side
is equal to zero only when  $\alpha(\varphi_{H})=0$, where  $\varphi_{H}$ is the value of the scalar field on the horizon,
which means that $\varphi_{H}=0$. We also require that $\varphi_{\infty}=0$, where $\varphi_{\infty}$ is the value of the scalar field at the
spacial infinity. In this situation, the only possibility to have a solution with non-trivial scalar field is
the scalar field to have at least one extremum. So let us integrate equation (\ref{phianl}) in the interval
$r\in[r_{H},r_{e}]$, where $r_{e}$ is the point of the leftmost (the one which is nearest to the event horizon) extremum of $\varphi$ which is on the right of the event horizon
\begin{eqnarray}
&&0=\left. \left(e^{-\delta}r^{2}f\frac{d\varphi }{dr} \right) \right|_{r=r_{H}}-
\left.\left(e^{-\delta}r^{2}f\frac{d\varphi }{dr} \right) \right|_{r=r_{e}} \nonumber \\
&&\hspace{2.3cm} =4\int\limits_{r_{e}}^{r_{H}}r^2 e^{-\delta} \alpha(\varphi)
{\cal A}^{4}(\varphi)\left[X\partial_{X}L(X)- L(X)\right] dr.\label{extrem}
\end{eqnarray}
Since $\varphi\neq0$ in the interval $(r_{H},r_{e}]$, the sign of $\alpha(\varphi)$ also does not change in this interval. Hence,
the integral on the
right-hand side of (\ref{extrem}) is non-zero and has a fixed sign which depends on the sign of $\alpha(\varphi)$.
The contradiction we reach means that no extremal solutions with non-trivial scalar field can exist.

To sum up, we can say that if a black hole exists it will have a
single horizon, i.e., its causal structure will be
Schwarzschild-like\footnote{Another class of scalar-tensor
theories which admit black holes of the Schwarzschild type are
those with negative function
$\beta(\varphi)={d\alpha(\varphi)\over d\varphi}$ for all values
of $\varphi$ and $\alpha(\varphi_{\infty})$=0. This can be shown
by  a method similar to that presented above. One can also show
that scalar-tensor theories with $\beta(\varphi)>0$ for all values
of $\varphi$ do not admit asymptotically flat black holes with
nontrivial scalar field.   }. In both conformal frames, inside the
event horizon a space-like singularity is hidden.

Finishing this section it is worth noting that the differential equations system (\ref{EQDelta}--\ref{EQPhi}) is invariant under the rigid rescaling
$r\to \lambda r$, $m \to \lambda m$, $P \to \lambda P$ and $b \to \lambda^{-2} b$ where $0<\lambda< \infty$. Therefore, given
a solution to (\ref{EQDelta}--\ref{EQPhi}) with one set of physical parameters $(r_{h}, M, P, b, {\cal D}, T)$, the rigid rescaling produces  new solutions with
parameters $(\lambda r_{h}, \lambda M, \lambda P, \lambda^{-2} b, \lambda {\cal D}, \lambda ^{-1}T)$. Here $T$ denotes the temperature of the horizon.

\section{Numerical results}
The nonlinear system (\ref{EQDelta})-(\ref{EQPhi}) is
inextricably coupled and the event horizon $r_H$ is {\it a priori} unknown boundary. In order to be solved numerically, it is
recast as a equivalent first order system of ordinary differential equations.  Following the physical assumptions of the
matter under consideration the asymptotic boundary conditions are set, i.e.,
$$\lim_{r \to \infty}m(r) =M \quad (M \>{\rm is\> the\> mass\> of\> the\> black\> hole\> in\> the\> Einstein\> frame}),$$
$$ \lim_{r \to \infty}\delta(r)=\lim_{r \to \infty}\varphi(r)=0.$$
At the horizon both the relationship $$f(r_H)=0$$
and
the regularization condition
$$\left.\left(\frac{df}{dr}\!\cdot\! \frac{d \varphi}{d r}\right)\right|_{r=r_H} =\left.
\left\{ 4 \alpha(\varphi) {\cal A}^4(\varphi) [X \partial_X L(X)-L(X) ]\right\}\right|_{r=r_H}$$ concerning the spectral quantity $r_H$ must be held.
For the treating the above posed boundary-value problem (BVP) the Continuous Analog of
Newton Method (see, for example \cite{gavurin},\cite{jidkov},\cite{YFBT}) is used. After an appropriate linearization the original BVP
is rendered to solving a vector two-point BVP. On a discrete level sparse (almost diagonal) linear algebraic
systems with regard to increments of sought functions $\delta(r)$, $m(r)$, and $\varphi(r)$ have to be inverted.

For our numerical solutions we have considered the STT studied by Damour and Esposito-Far\`{e}se in their works on neutron
stars in the STT of gravity. In this particular theory, the
coupling function has the following form
\begin{equation}
{\cal A}(\varphi)=e^{\frac{1}{2}\beta\varphi^2}, \\
\end{equation}
where $\beta$ is a negative constant. Observational data from binary-pulsar and solar-system experiments
restricts the admissible values of $\beta$.
When $\alpha(\varphi=0)=0$, as it is in our case, the coupling constant should be $\beta>-5$
(see \cite{DF2} for more details).
We have studied the solutions for values of the parameters which are in agreement with the current observations but also
for such values that are out the admissible interval since the later have qualitatively different behavior from the
former which makes them interesting for the theory.

For this coupling function the field equations possess the discrete symmetry
$\varphi \to -\varphi$. Let us also note that every general relativistic solution is a solution to the
scalar tensor theory under consideration with $\varphi=0$.

A thorough study of the phase space would be difficult due to the large number of parameters ($\beta, b, P, M$) in the problem.
So in order to illustrate the general behavior of the obtained solutions we
give several representative figures considering several values of $\beta, b$ and $P$ and varying the mass $M$ of the black hole. For the cases presented here
$b=0.01; \, 0.2 $.

\subsection{$\textbf{General description of the phase space}$}
Even if the values of all four parameters $\beta, b, P, M$ and the boundary conditions are fixed the solutions of (\ref{EQDelta}-\ref{EQPhi}) are not uniquely determined, i.e. the solutions are not unique. An additional parameter should be used for labeling of the different solutions. One natural choice\footnote{An alternative choice would be the value of the scalar field on the horizon } would be to label the different solution by the value of the scalar charge ${\cal D}$, which here unlike the cases in \cite{SYT1, SYT2} is independent.

\begin{figure}[htbp]%
\vbox{ \hfil \scalebox{1.0}{ {\includegraphics{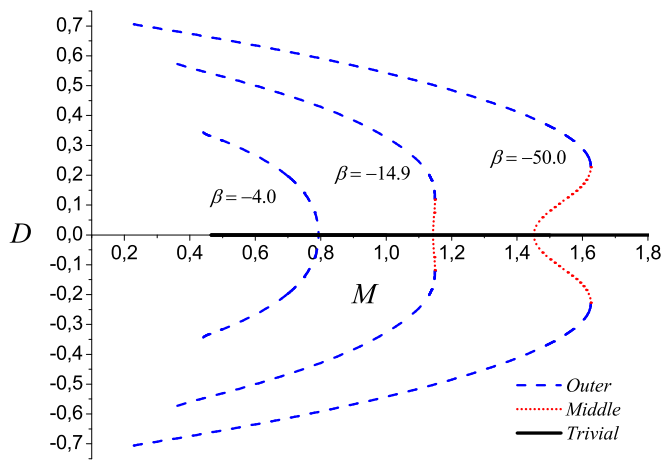}} }\hfil}%
\bigskip%
\caption{%
The $M-{\cal D}$ relation for several different values of $\beta$, $b=0.01$ and $P=1.0$. For $\beta>\beta_{\rm {crit}}$ the \emph{Middle} branch disappears.} \label{beta_crit}%
\end{figure}

The global structure of the phase space changes with the variation of $\beta$. For $\beta>\beta_{\rm {crit}}$, where $\beta_{\rm {crit}}\approx-14.9$ when $b=0.01$ and $P=1.0$, the $M-{\cal D}$ phase diagram consists of three branches, while for $\beta<\beta_{\rm {crit}}$ the number of branches is five. The change of the qualitative structure of that phase diagram with the variation of $\beta$ is given in Figure (\ref{beta_crit}). The different cases have been studied in more details in the succeeding subsections.

\subsection{$\textbf{Cases with three solutions, $\beta=-4.0$}$}

\begin{figure}[htbp]%
\includegraphics[width=0.54\textwidth]{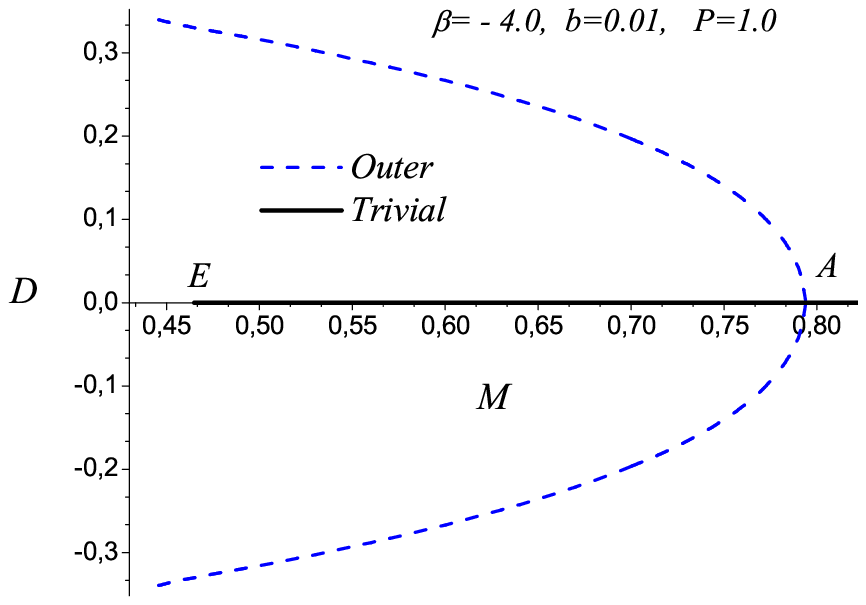} 
\includegraphics[width=0.49\textwidth]{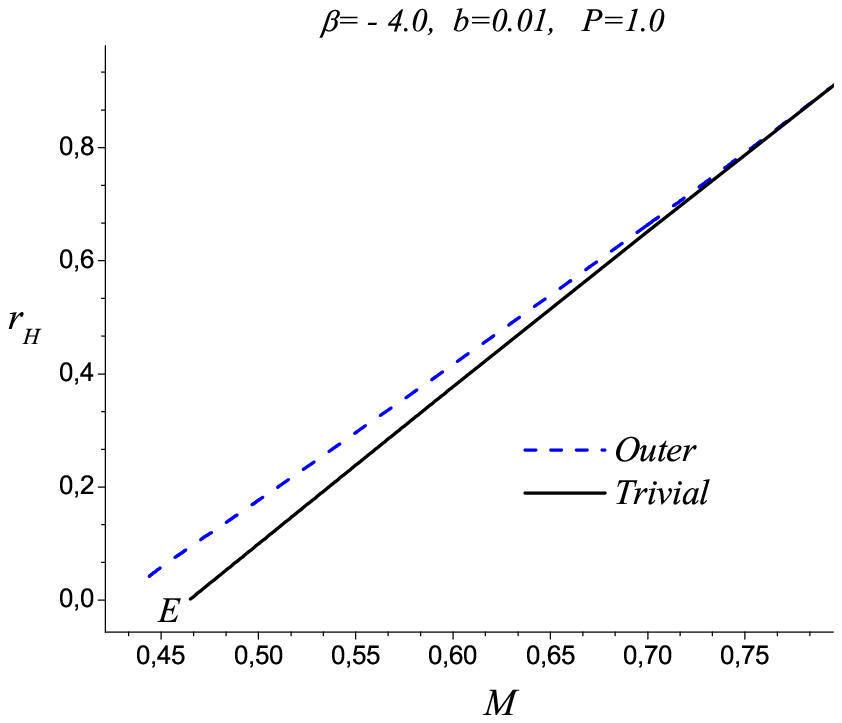}
\caption{%
The $M-{\cal D}$ and the $M-r_{H}$ relations, for $\beta=-4.0$, $b=0.01$ and $P=1.0$.} \label{DM_Rh_4}%
\end{figure}%

\begin{figure}[htbp]%
\vbox{ \hfil \scalebox{0.9}{ {\includegraphics{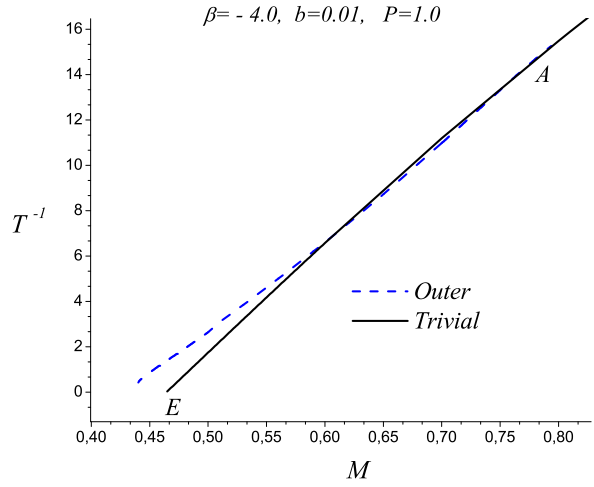}} }\hfil}%
\bigskip%
\caption{%
The $M-T^{-1}$ relation for the same values of the magnetic charge as in Figure (\ref{DM_Rh_4}).} \label{TM_4}%
\end{figure}%

In the left panel of Figure (\ref{DM_Rh_4}) the dependence ${\cal D}(M)$ of the scalar charge on the mass for $\beta=-4.0$ is presented. There are two special points to be considered, namely $A$ and $E$. For $P=1.0$ the points $A$ and $E$ lie at masses  $M_{A}\approx0.794$ and $M_{E}\approx0.46$.
For values of the black hole mass in the interval $M\in(M_{E},M_{A})$
three solutions co-exist. We call them \emph{Outer} and \emph{Trivial}.
The equations posses a discrete symmetry $\varphi\rightarrow-\varphi$ as a result of which the \emph{Outer} solution
has a mirror image with respect to the abscissa, whose scalar field charge ${\cal{D}}$ has an opposite sign.
The symmetric solutions we will denote as $Outer_{\pm}$ where the indices $+$ and $-$ refer to
the sign of the scalar field charge of the solutions. The radius and the temperature of the event horizon, are the same for both solutions in the couple so when we comment on them will usually omit the $+$ and $-$ indices.

The \emph{Trivial} brunch has scalar field
$\varphi\equiv0$ and represents the Einstein-Born-Infeld solution in GR.
For masses lower than $M_{E}$ only the solutions which we call \emph{Outer} exist and for
$M>M_{A}$ only the \emph{Trivial} solution remains.

The radii of the black holes are shown in the right panel of Figure (\ref{DM_Rh_4}). An object with zero radius of the
event horizon (a naked singularity) is reached for a finite value of $M$ for all three solutions -
the $Outer_{\pm}$ and the \emph{Trivial}.

The inverse temperature $T^{-1}$ of the solutions is presented in Figure (\ref{TM_4}).

\subsection{$\textbf{Cases with five solutions, $\beta=-50.0$}$}

\subsubsection{$\textbf{Solutions with Born-Infeld parameter b=0.01}$}

\begin{figure}[htbp]%
\includegraphics[width=0.50\textwidth]{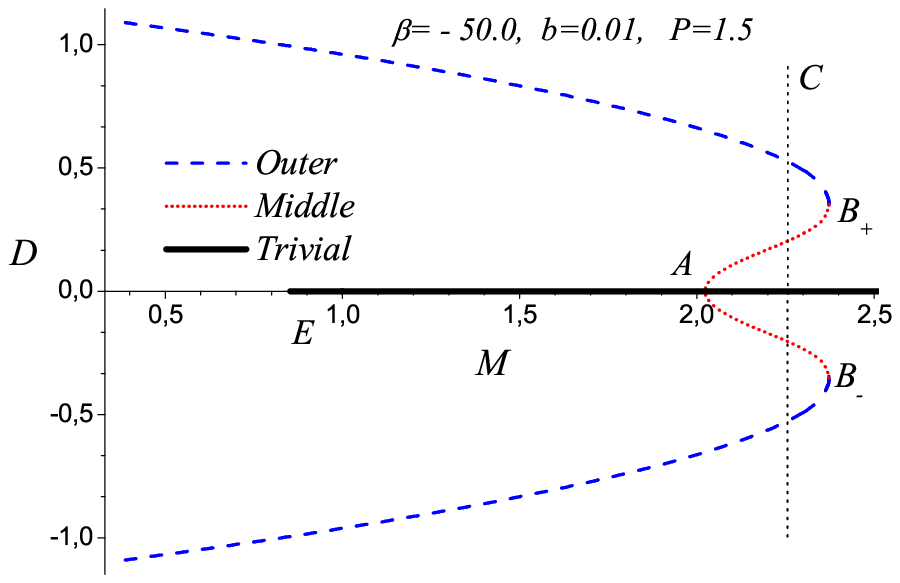}
\includegraphics[width=0.47\textwidth]{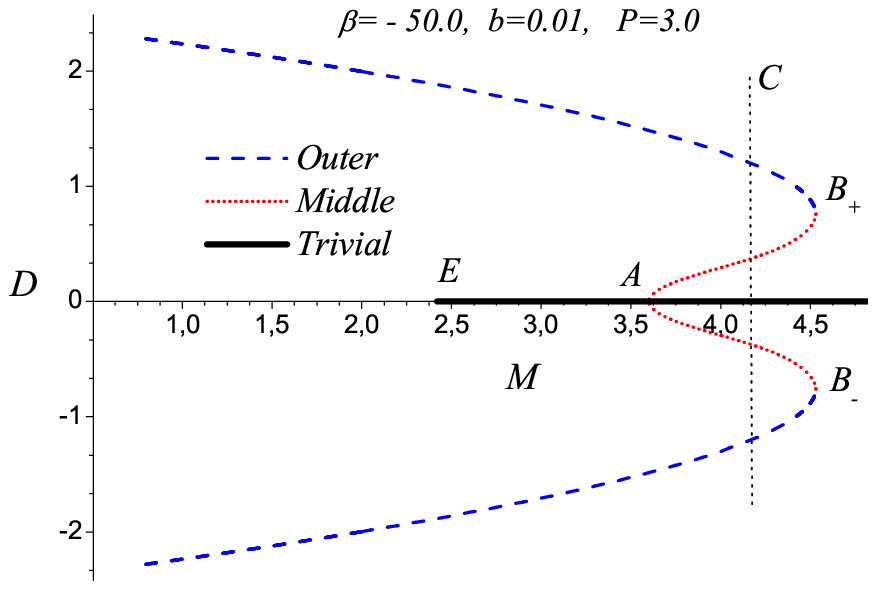}
\caption{%
The value of scalar field charge ${\cal D}$ as a function of the mass $M$ of the black hole, for
two different values of the magnetic charge $P=1.5$ and $P=3.0$.} \label{DM}%
\end{figure}%

\begin{figure}[htbp]%
\includegraphics[width=0.50\textwidth]{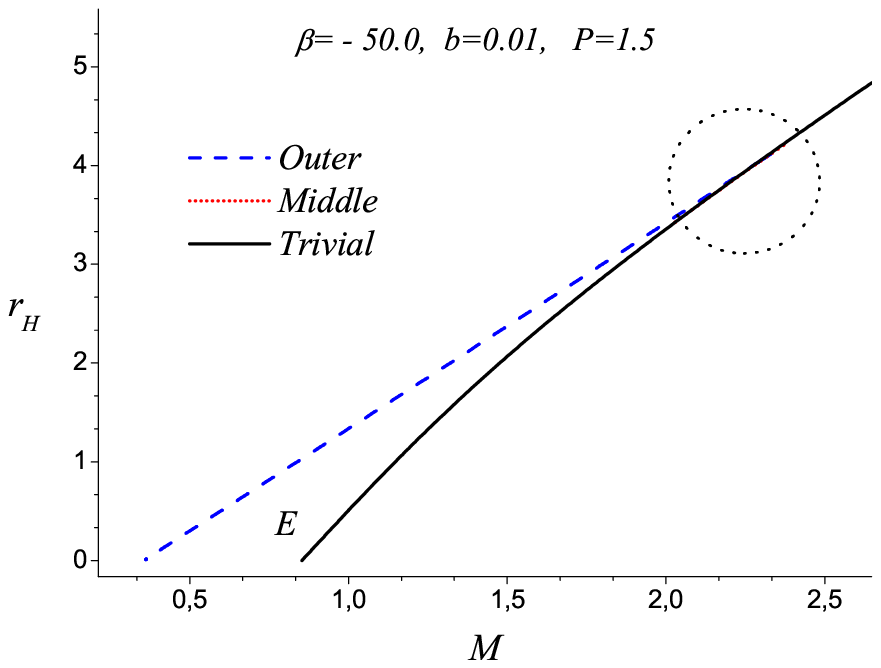}
\includegraphics[width=0.50\textwidth]{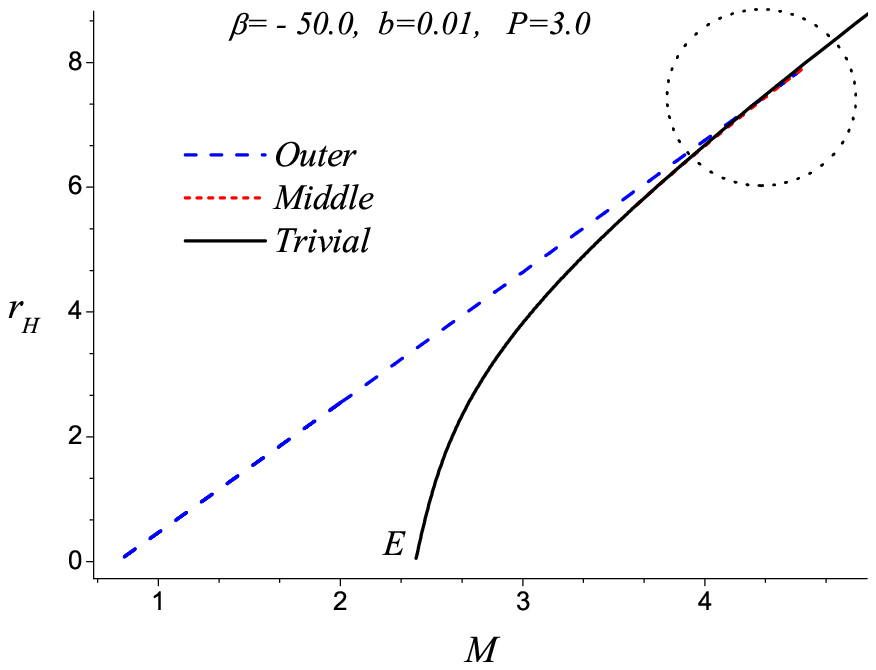}
\caption{%
The $M-r_{H}$ relation for the same values of the magnetic charge as in Figure (\ref{DM}).
A magnification in the encircled region is presented in Figure (\ref{RhM_Uvel}).} \label{RhM}%
\end{figure}%

\begin{figure}[htbp]%
\includegraphics[width=0.50\textwidth]{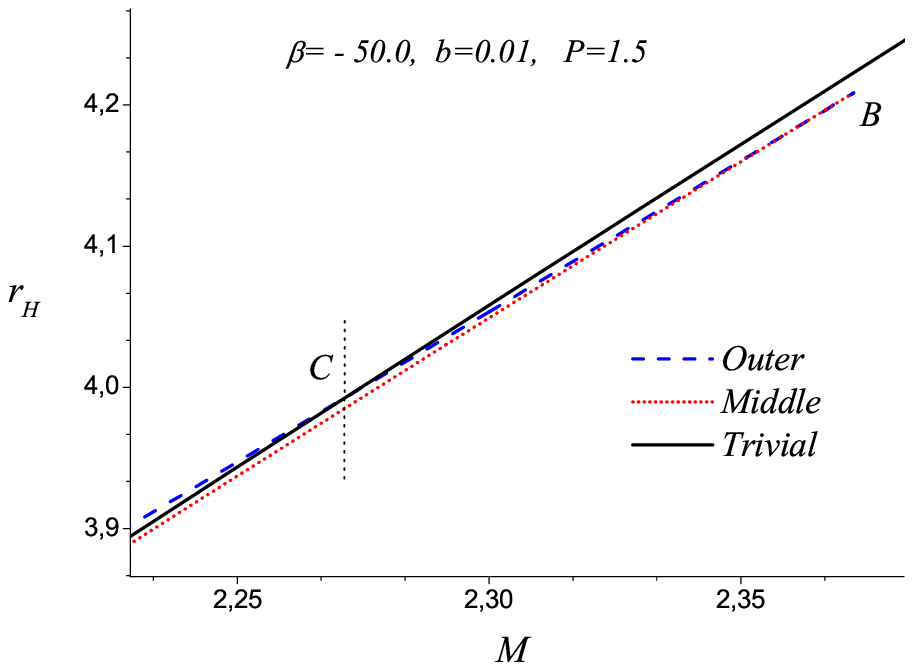}
\includegraphics[width=0.50\textwidth]{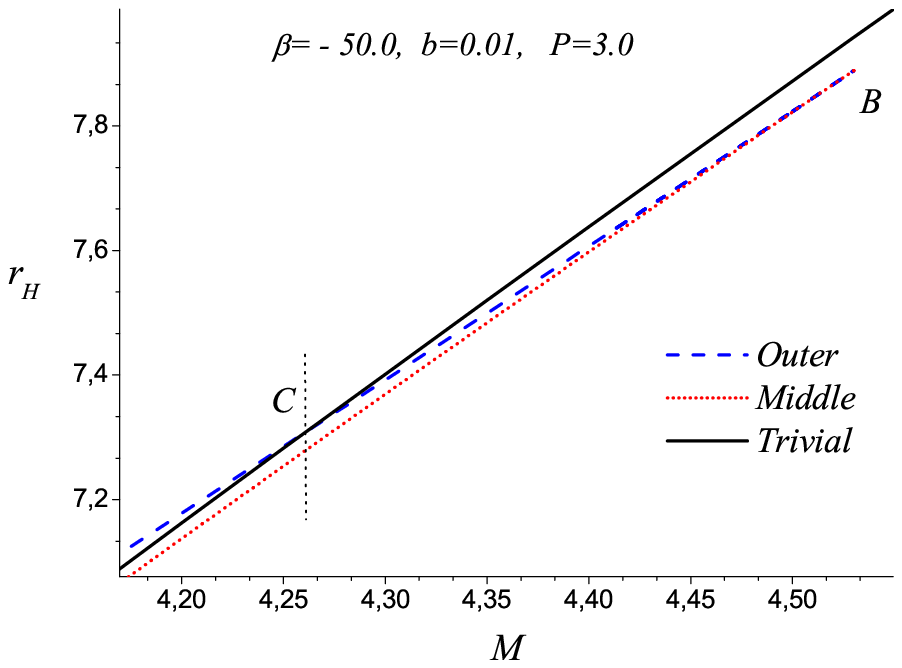}
\caption{%
A magnification of the encircled region in Figure (\ref{RhM}).} \label{RhM_Uvel}%
\end{figure}%

\begin{figure}[htbp]%
\includegraphics[width=0.50\textwidth]{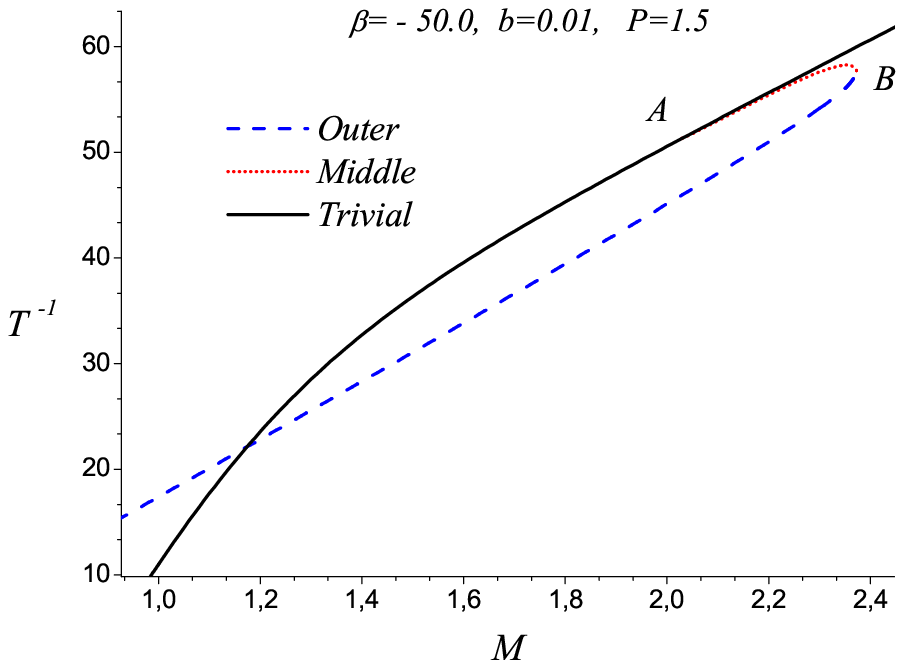}
\includegraphics[width=0.50\textwidth]{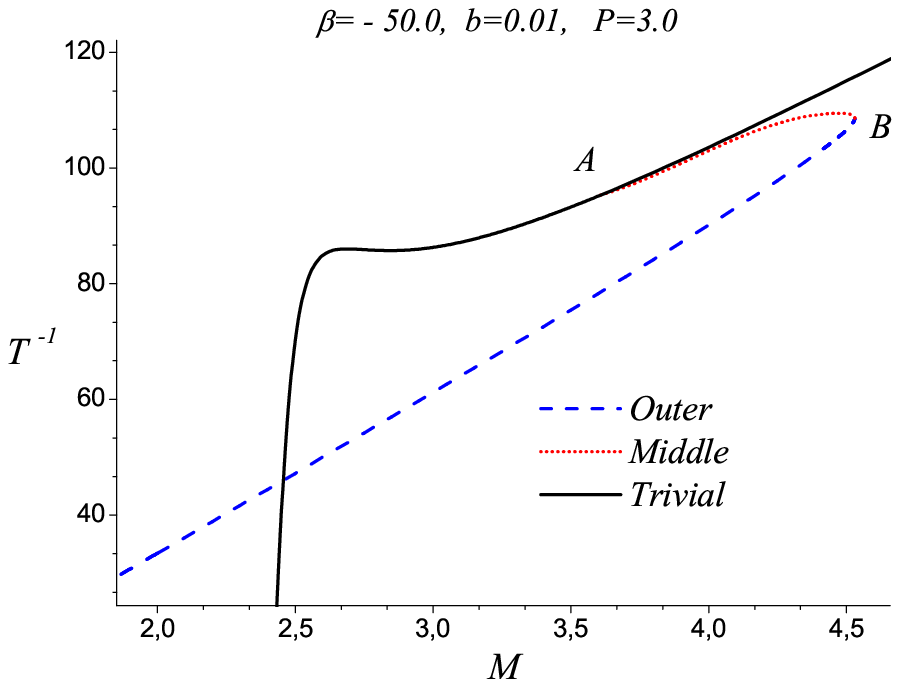}
\caption{%
The $M-T^{-1}$ relation for the same values of the magnetic charge as in Figure (\ref{DM})}. \label{TM}%
\end{figure}%


For $\beta=-50.0$ and $b=0.01$ the ${\cal D}(M)$ dependence is given in Figure (\ref{DM}). Here, the special points to be considered are four, $A$, $B_{\pm}$
and $E$. For $P=1.5$ the points $A$ and $B_{\pm}$ lie at masses  $M_{A}\approx2.023$ and  $M_{B_{\pm}}=M_{B}\approx2.37$ while in the case of $P=3.0$
we have  $M_{A}\approx3.60$ and  $M_{B_{\pm}}=M_{B}\approx4.53$.
As it can be seen, for values of the black hole mass in the interval $M\in(M_{A},M_{B})$
five solutions co-exist. They
can be separated in three groups which
we name \emph{Outer}, \emph{Middle} and \emph{Trivial}.
The symmetric solutions we will denote as $Outer_{\pm}$  and $Middle_{\pm}$ with the same convention for the $+$ and $-$ indices as in the previous case. Both points $B_{+}$ and $B_{-}$ are
projected on the same point on the $M-r_{H}$ and $M-T^{-1}$ diagrams, which we denote simply as $B$.

The \emph{Trivial} has scalar field
$\varphi\equiv0$ and represents the Einstein-Born-Infeld solution in GR.
For masses lower than $M_{E}\approx1.00$ and $M_{E}\approx2.42$ for $P=1.5$ and $P=3.0$, respectively,  only the solutions which
we call \emph{Outer} exist and for
$M>M_{B}$ only the \emph{Trivial} solution remains.

The radii of the black holes are shown in Figures (\ref{RhM}) and (\ref{RhM_Uvel}). Again, for three of the solutions -
the \emph{Outer} and the \emph{Trivial} a naked singularity is reached for a finite value of $M$. The \emph{Outer} solutions have a larger radius than the
\emph{Trivial} for low masses, but as it can be seen in Figure (\ref{RhM_Uvel}), which is a
magnification of the encircled region in Figure (\ref{RhM}), with the increase of the mass, in point $C$,
the situation changes and the
black hole with zero scalar charge becomes larger. The approximate position of point $C$ is shown also on
Figure (\ref{DM}) with a dotted vertical line. The \emph{Middle} solution black holes are smaller than the other
three for all values of $M$.

The inverse temperature $T^{-1}$ of the solutions is presented in Figure (\ref{TM}). With the
decrease of the mass $M$ the inverse temperature of the \emph{Trivial} solution passes through
a local maximum which gets sharper with the increase of the magnetic charge $P$ and
leads to numerical calculation difficulties. In the limit of vanishing radii of the black holes
their inverse temperature decreases steeply.

\subsubsection{$\textbf{Solutions with Born-Infeld parameter b=0.2}$}

\begin{figure}[htbp]%
\includegraphics[width=0.53\textwidth]{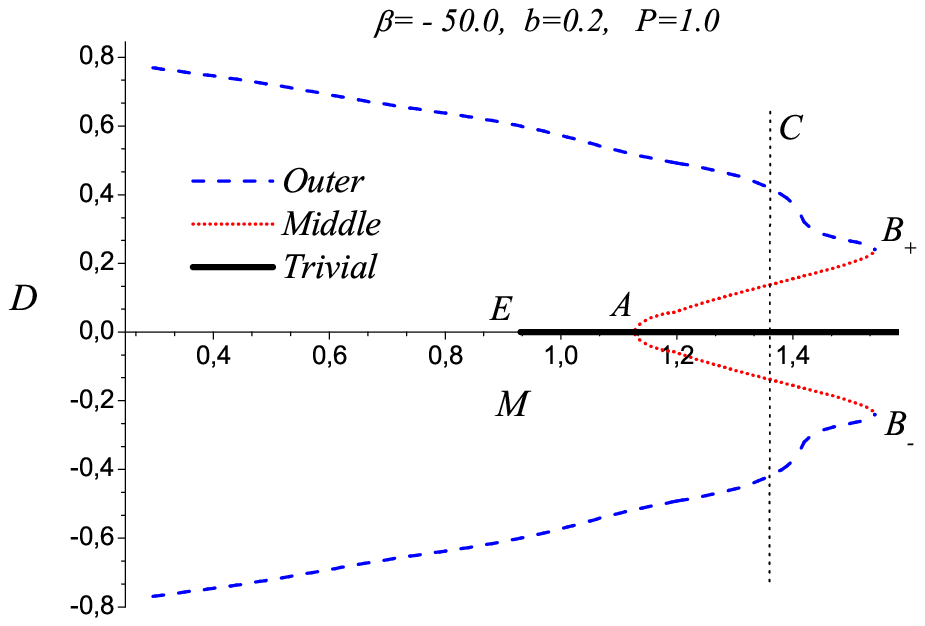} 
\includegraphics[width=0.47\textwidth]{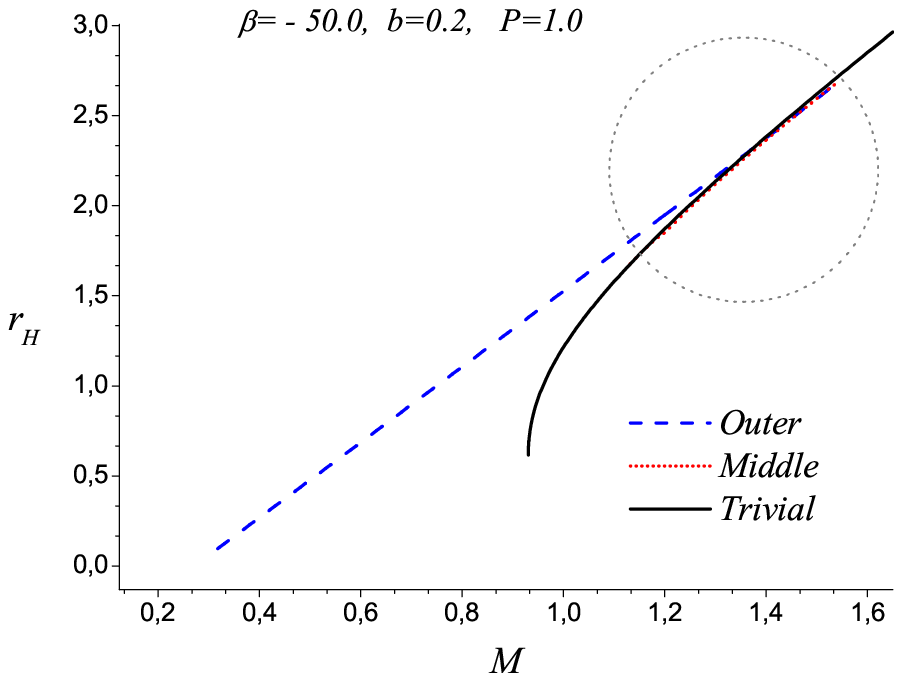}
\caption{%
The $M-{\cal D}$ and the $M-r_{H}$ relations, for $b=0.2$ and $P=1.0$.} \label{DM_Rh}%
\end{figure}%

\begin{figure}[htbp]%
\includegraphics[width=0.47\textwidth]{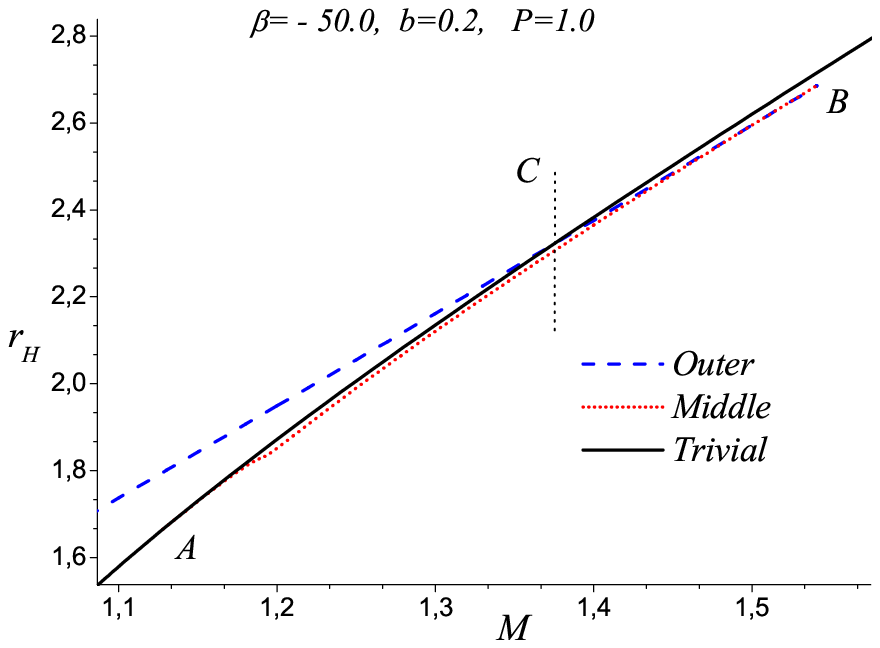}
\includegraphics[width=0.50\textwidth]{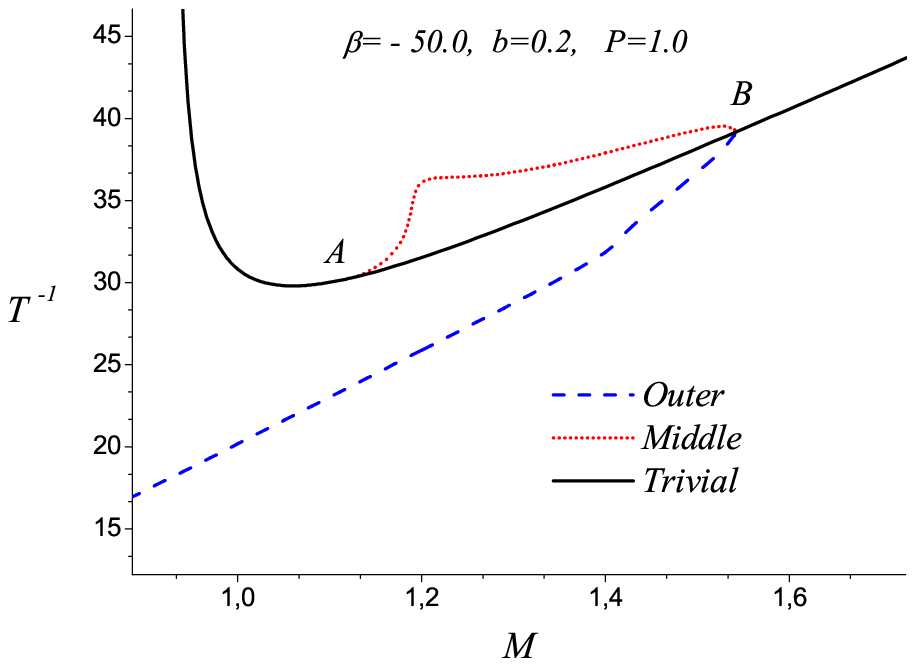}
\caption{%
A magnification in the encircled region of the $M-r_{H}$ relation from Figure (\ref{DM_Rh}) and the $M-T^{-1}$ relation, for $b=0.2$ and $P=1.0$.
.} \label{RhM_TM}%
\end{figure}%

Here we present an example of solutions for values of the parameters for which the \emph{Trivial} solution reaches an extremal black hole
instead of a naked singularity with the decrease of the mass $M$.
In Figure (\ref{DM_Rh}) the dependence ${\cal D}(M)$ of the scalar charge on the mass and the radii of the black holes are presented.
The points $A$ and $B_{\pm}$ lie at masses  $M_{A}\approx1.13$ and  $M_{B_{\pm}}=M_{B}\approx1.54$.
Again, for values of the black hole mass in the interval $M\in(M_{A},M_{B})$
five solutions co-exist. The \emph{Trivial} reaches an extremal black hole at $M_{E}\approx0.93$ and this can be seen on the $M-r_{H}$ diagram.

In Figure (\ref{RhM_TM}) a magnification of the encircled region from Figure (\ref{DM_Rh}) and the inverse temperature
are shown.  The point $C$ is once again indicated with a vertical line.
With the approaching of the extremal black hole the inverse temperature $T^{-1}$ rises unboundedly.

\section{Thermodynamics}
Black holes have long been known to be thermodynamical systems \cite{Beken}.
The First Law (FL) of black hole thermodynamics in the presence of a scalar field has the following form \cite{Rasheed}
\begin{equation}
\delta M=T \delta S + \Psi_{H}\delta P+{\cal{D}}\delta \varphi_{\infty},\label{FirstLaw}
\end{equation}
where $T$, $S$ and $P$ are the temperature, the entropy, and the magnetic charge of the black hole, respectively, and
$\varphi_{\infty}$ is the value of the scalar field at spacial infinity.

The quantity $\Psi$ conjugate to the magnetic charge  is the potential of the magnetic field which is given by the
following definition
\begin{equation}
H_{\mu}=\partial_{\mu}\Psi.\label{defH}
\end{equation}
On the other hand the magnetic field is defined as
\begin{equation}
H_{\mu}=-\star G_{\mu\nu}\xi^{\nu},
\end{equation}
where
\begin{equation}
G_{\mu\nu}=- 2 \frac{\partial \left({\cal A}^{4}(\varphi)L\right)}{\partial F_{\mu\nu} },
\end{equation}
$\xi=\frac{\partial}{\partial t}$ is the Killing vector generating time translations
and ``$\star$'' is the Hodge star operator.

Since in our case the asymptotic value of scalar field is
fixed, the term which contains its variation vanishes and the FL reduces
to the form it has in GR.

In the situation when the solutions are not unique a natural question is which of them are stable. Certainly,
the answer of this question requires linear perturbative analysis of our system of coupled  differential
equations and solving the corresponding eigenvalue problem. In certain  cases, however,  some information  on the stability
can be inferred by using only the equilibrium thermodynamical characteristics of the solutions via the so-called ``Poincare" or
``turning point" method. For a nice discussion of the method we refer the reader to  \cite{Arcioni} and references therein. The advantage
of this method is its remarkable simplicity. The ``turning point" method, however, may  hide many uncertainties
and should be applied with caution.
The method consists in the following. Consider a system with thermodynamical parameters $\mu^{i}$. The equilibrium states (stable or not)
are extrema of an appropriate  Massieu function ${\cal S}$. At equilibrium state we can define the conjugate variables $\beta_{i}(\mu^j)=
{\partial {\cal S}_{eq}\over \partial \mu^{i} }$.
According to the method the change of stability can only  occur at turning points\footnote{The turning points are points
where two equilibrium branches merge with a vertical tangent. A bifurcation point is point where branching of equilibrium sequences occurs.}
or bifurcations in the equilibrium sequence on the conjugate diagram $\beta_{i}(\mu^{i})$. In the absence of bifurcations
the stability character changes only when
the equilibrium curve meets a turning point and if one single point of an equilibrium sequence is shown to be fully stable,
then all equilibria in the sequence are fully stable up to the first turning point. At the turning point the branch with
negative slope is always unstable while the branch with a positive slope is more stable than the one with negative slope.

Concerning the application of the ``turning point" method  to our case we shall consider scalar-tensor black  holes in the
micro-canonical ensemble. In this case the corresponding Massieu function is the entropy $S(M)$\footnote{We keep the magnetic charge fixed.}.
We will comment on the case with five branches first. The conjugate
variable is $\beta_{M}=T^{-1}$. For $\beta<\beta_{\rm {crit}}$ the conjugate diagram $M-T^{-1}$ is shown on Figures (\ref{TM}) and (\ref{RhM_TM}). The uncertainties in our case come
from that fact that we do not know the full diagram, i.e., whether there are other branches and bifurcation points different from
point $A$. Assuming however that there are no other bifurcation points and taking into account that there is only one
turning point $B$ we may conclude that the \emph{Middle} branch is unstable
while the  \emph{Outer} branch is more stable. The cusp which appears in point $B$
on the $M-r_{H}$ diagram in Figures (\ref{RhM_Uvel}) and (\ref{RhM_TM}) also indicates a change in the stability of the solutions.
Since the \emph{Outer} branch is the unique solution (up to the discrete degeneracy  $\emph{Outer}_{\pm}$) for sufficiently small masses
one might accept that the \emph{Outer}  branch is probably stable there. Then according to the  ``turning point" method
the  \emph{Outer} branch should be  fully stable  up to the turning point $B$. The \emph{Trivial} branch is unique solution for
sufficiently large masses and as a GR solution is known to be stable\footnote{In general, the stability of the trivial solution
within the framework of GR does not guarantee  its stability within the ``larger" scalar-tensor theory.} there \cite{Breton, Fernando}. So, assuming that \emph{Trivial} branch  is stable for
large masses, the  ``turning point" method
asserts that the \emph{Trivial} branch should be stable up to the bifurcation point $A$.
Since the entropy of a black hole is proportional to the area of its event horizon, among the three black-hole
phases we consider, the one with the biggest radius would have the maximal entropy (see Figures (\ref{RhM}), (\ref{RhM_Uvel}), (\ref{DM_Rh}) and (\ref{RhM_TM})).
So for $M<M_{C}$ the \emph{Outer} solutions would be thermodynamically favorable and
for $M>M_{C}$ - the \emph{Trivial}.
The \emph{Middle} solutions are  thermodynamically unstable for all values of the mass since their radius is
smaller than the radii of the other three solutions. The point $C$ is a candidate for a point of a first order phase transition.
Let us stress again that above analysis based on the ``turning point" method is only suggestive
and cannot serve for a definitive solution of the stability of the solutions.
Reliable analysis of the risen questions will be given elsewhere together with the solution of corresponding
eigenvalue problem.

For $\beta>\beta_{\rm {crit}}$ the $M-T^{-1}$ diagram is given in  Figure (\ref{TM_4}). No turning points can be
seen on it. As in the previously discussed case, since $\emph{Outer}_{\pm}$ are the only solutions for sufficiently
small masses one might accept that they are probably stable there. Then according to the ``turning point" method
the  \emph{Outer} branch should be fully stable up to the bifurcation point $A$. Again, assuming that \emph{Trivial}
branch is stable for large enough masses one can expect that to the left it should be stable at least up to the bifurcation
point $A$.
From the right panel of Figure (\ref{DM_Rh_4}) we can see that the \emph{Outer} solutions have larger radius of the
event horizon so they would be thermodynamically favorable.

The thermodynamical stability considerations were made in the Einstein frame. In order to transfer the
conclusions to the physical, Jordan frame properly, we need to clarify the connection between the
thermodynamic properties in the two conformal frames.
The temperature of the event horizon is invariant under
conformal transformations of the metric that are unity at infinity \cite{Jacobson}.
The properly defined entropy is also the same in both conformal frames.
It has been proved that in the Jordan frame the entropy of the black hole is not simply one fourth of
the horizon area \cite{MVisser,FordRoman} as in the Einstein frame and needs to be generalized. The entropy in the
Jordan frame is defined
as
\begin{equation}
S_{J}=\frac{1}{4G_{*}}\int d^2x \sqrt{-^{(2)}{\tilde g}}F(\Phi).
\end{equation}
Passing to the Einstein frame we get
\begin{equation}
S_{J}=\frac{1}{4G_{*}}\int d^2x \sqrt{-^{(2)} g}=S_{E}=S.
\end{equation}
In the last two equations $^{(2)}{\tilde g}$ and $^{(2)} g$ are the determinants of the induced metrics on the horizon
in the Jordan
and in the Einstein frame, respectively.

The term in the FL (\ref{FirstLaw}) connected with the magnetic charge is also preserved under the conformal
transformations.

In order for the FL of thermodynamics to be satisfied in the Jordan frame the mass should be
properly chosen since
the Arnowitt-Deser-Misner (ADM) masses in both frames are not equivalent. It can be easily shown that in the STT considered the
ADM mass in the Jordan frame $M_{J}$ is equal
ADM mass in the Einstein frame $M$
\begin{equation}
M_{J}=M.
\end{equation}
For the Jordan frame, the proper mass in the FL of thermodynamics is the ADM mass in the Einstein
frame $M$. Similarly, for boson and fermion stars the proper measure for the energy of the system is again the
ADM mass in the Einstein frame $M$. For more details on the subject
we would refer the reader to the works \cite{Lee,Shapiro,Whinnett,Yazadji}.

\section{Conclusion}

In the present work new numerical solutions describing multiple-phase, charged black holes coupled to
non-linear electrodynamics in
the scalar-tensor theories with massless scalar field were found. Since an electric-magnetic duality is present,
here only the purely magnetically charged case was studied. For the Lagrangian of the non-linear electrodynamics the
truncated Born-Infeld
Lagrangian was chosen and a special STT was considered. As a result of the
numerical and analytical investigations, some general properties of the solutions were found.
In the class of STT for which $\varphi\,\alpha(\varphi)\leq0$ for all values of $\varphi$,
black holes with a single horizon exist, i.e., their causal
structure resembles that of the Schwarzschild black hole and is simpler than the corresponding solution in GR.
In dependance of the black hole mass, different black hole phases, labeled by the scalar charge coexist. For sufficiently small masses
there are two phases and only one phase for sufficiently large masses. In the intermediate range of black hole mass, there are
three or five coexisting phases.

The thermodynamical stability of the solutions was also discussed but the present data is not
sufficient for a final conclusion. The full understanding of the stability of the phases and the dynamics of possible phase transitions requires
perturbative analysis.

Presence of multiple phases is also expected in cases of other sources of gravity whose tensor of energy-momentum has a non-vanishing trace,
such as: linear and non-linear electrodynamics in higher dimensions (or even in 3D gravity), gauge theories and so on.

\section*{Acknowledgments}
We would like to thank G.~Arcioni, G. Esposito-Far\`{e}se and B.~Kol for the correspondence and valuable
comments, and to N. Karchev, T. Mishonov and P. Nikolov for the discussions on the
multiple phases and phase transitions.
This work was partially supported by
the Bulgarian National Science Fund under Grants MUF 04/05 (MU 408), VUF-201/06 and the
Sofia University Research Fund N111. It was supported in part also by the European
Research and Training Network ``Constituents, Fundamental Forces and
Symmetries of the Universe'' under contract number
MRTN-CT-2004-005104. S.Y. would like to thank the Alexander von
Humboldt Foundation for a stipend, and the Institut f\"{u}r
Theoretische Physik G\"{o}ttingen for its kind hospitality. I.S.
would like to thank the University of Patras for the kind
hospitality. M.T. is very grateful to the Department of Mathematics of the University of Texas at
Arlington, USA for the invitation and the kind hospitality during Spring
semester 2008.

\end{document}